\documentstyle[twocolumn,aps,pra,epsf]{revtex}
\begin{document}
\draft
\title{Nonlocality without inequalities has not been proved\\
for maximally entangled states\thanks{To appear in Phys. Rev. A.}}
\author{Ad\'{a}n Cabello\thanks{Electronic address:
fite1z1@sis.ucm.es}}
\address{Departamento de F\'{\i}sica Aplicada,
Universidad de Sevilla, 41012 Sevilla, Spain}
\date{\today}
\maketitle
\begin{abstract}
Two approaches to extend Hardy's proof
of nonlocality without inequalities to maximally entangled states
of bipartite two-level systems
are shown to fail. On one hand,
it is shown that Wu and co-workers' proof
[Phys. Rev. A {\bf 53}, R1927 (1996)] uses
an effective state which is not maximally entangled.
On the other hand, it is demonstrated
that Hardy's proof cannot be generalized by the replacement of one
of the four von Neumann measurements involved in the original
proof
by a generalized measurement to unambiguously discriminate between
non-orthogonal states.
\end{abstract}

\narrowtext

\section{Introduction}
Hardy's proof of ``nonlocality without inequalities'' \cite{Hardy93}
provides the simplest demonstration of Bell's theorem \cite{Bell64}
that there is no local realistic theory reproducing all
predictions of quantum mechanics.
Curiously, while the maximum violation of Bell inequalities
occurs for maximally entangled states \cite{Kar95},
Hardy's proof does not go through for maximally entangled states.
Recently, Wu, Xie, Huang, and Hsia (WXHH) \cite{WXHH96} have
claimed to have demonstrated nonlocality without
inequalities for bipartite two-level
systems prepared in a maximally entangled state.
Their approach
is based on a selection of events in a modified version
of the two-particle interferometer proposed by Horne, Shimony, and
Zeilinger \cite{HSZ89}.
In Sec.\ III of this paper WXHH's approach is reexamined.
I will argue that it fails to
prove nonlocality without inequalities for maximally entangled states.
After this analysis, it will become clear that no extension
of Hardy's proof to maximally entangled states is possible by selecting
events before the local measurements involved in the
proof.
Therefore, it would be interesting to investigate whether such an extension
can be achieved by using a set-up in which
the selection of events necessarily occurs after the local
measurements. In particular, I will investigate whether Hardy's
proof can be generalized by the replacement of
one of the four von Neumann local
measurements
by a measurement to unambiguously discriminate between
non-orthogonal states \cite{Ivanovic87,Dieks87,Peres88,Peres93}.
This scenario was considered for a different purpose by Chefles
and Barnett \cite{CB97}. In Sec.\ IV a
general demonstration showing that Hardy's proof cannot be generalized
in such a way will be provided.
Our discussion begins in
Sec.\ II, where Hardy's \cite{Hardy93} and Goldstein's \cite{Goldstein94}
versions of Hardy's proof are reviewed.
By ``versions'' I mean logical reasonings based on the same
set of properties of certain quantum states.
This distinction between versions
will be useful in Sec.\ IV.

\section{Nonlocality without inequalities
for Hardy states}
We shall focus our attention on
bipartite two-level systems initially
prepared in a state of the form
\begin{equation}
\left| \psi \right\rangle =
a\left| ++ \right\rangle + b
\left( {\left| +- \right\rangle+
\left| -+ \right\rangle} \right),
\label{H1}
\end{equation}
where $a=\cos\theta$, and $b=\sin\theta / \sqrt 2$,
being $0\le\theta\le \pi / 2$.
The notation
$\left| {+-} \right\rangle$ means
$\left| + \right\rangle _1\otimes \left| - \right\rangle _2$,
being $\left\{ \left| + \right\rangle_j,
\left| - \right\rangle_j \right\}$
an orthogonal basis for particle $j$ ($j=1,2$).

Now I shall explain why the study
of the family of states given by Eq.\ (\ref{H1}) covers all relevant cases.
For bipartite pure states, partial entropy is a good measure of
entanglement \cite{BBPS96,LP97} since it fulfills the following requirements
\cite{VPRK97}: to have zero value for product states,
to be invariant under local unitary transformations
and non-increasing under classically coordinated local
operations, and to be additive for tensor products.
From the properties of partial entropy, it follows that
any two pure states having the same partial entropy will give the same
maximum probability for finding an event which contradicts local realism for a
standard Hardy's proof.
Therefore, the conclusions reached for a state of the form (\ref{H1})
with partial entropy $S$, can be extended to any bipartite two-level pure
state with partial entropy $S$.
Partial entropy of states of the form (\ref{H1}) is
a monotone function of the angle $\theta$,
and takes the value zero, for $\theta = 0$, and
the maximum allowed value,
$\ln 2\approx 0.6931$,
for $\theta = \pi / 2$.
Therefore, states of the form (\ref{H1}) cover
all possible values of partial entropy
and thus they cover all possible cases of
contradiction with local realism.
Moreover, this
partial entropy depends on a single parameter $\theta$:
If $\theta = 0$, then $\left| \psi \right\rangle$ is
a product state;
if $0 < \theta < \pi / 2$, then
$\left| \psi \right\rangle$ is an entangled
but not maximally entangled state;
and if $\theta = \pi / 2$, then
$\left| \psi \right\rangle$ is a maximally entangled state.

Suppose $ \left| + \right\rangle_j$ and
$\left| - \right\rangle_j$ are the eigenstates corresponding to
the observable $A_j$, and
$ \left| \oplus \right\rangle_j$ and
$\left| \ominus \right\rangle_j$ are the eigenstates corresponding to
the observable $B_j$, being
\begin{mathletters}
\begin{eqnarray}
\left| \oplus \right\rangle _j
=N \left( {a\left| + \right\rangle _j+
b\left| - \right\rangle _j} \right), \\
\left| \ominus \right\rangle _j
=N \left( {b\left| + \right\rangle _j-
a\left| - \right\rangle _j} \right),
\end{eqnarray}
\end{mathletters}
where $j=1,2$, and $N =1 / \sqrt {1-b^2}$.
Then state (\ref{H1}) can be written
in the following forms:
\begin{mathletters}
\begin{eqnarray}
\left| \psi \right\rangle & = &
N \left[ {\left( {1-b^2} \right)\left| {\oplus +} \right\rangle +
ab\left| {\oplus -} \right\rangle +
b^2\left| {\ominus -} \right\rangle } \right], \label{H2} \\
& = & N \left[ {\left( {1-b^2} \right)\left| {+ \oplus} \right\rangle +
ab\left| {- \oplus} \right\rangle +
b^2\left| {- \ominus} \right\rangle } \right]. \label{H3}
\end{eqnarray}
\end{mathletters}
Now, we distinguish
between two versions of the proof.

\subsection*{Hardy's proof}
From Eqs. (\ref{H2}), ({\ref{H3}), and ({\ref{H1}), respectively,
it can easily be seen that any state $\left| \psi \right\rangle$
has the following properties:
\begin{mathletters}
\begin{eqnarray}
P_\psi \left( {\left. -_2\, \right|\, \ominus_1} \right) & = & 1,
\label{Har1} \\
P_\psi \left( {\left. -_1\, \right|\, \ominus_2} \right) & = & 1,
\label{Har2} \\
P_\psi \left( -_1,-_2 \right) & = & 0.
\label{Har4}
\end{eqnarray}
In addition, as can be easily checked,
\begin{equation}
P_\psi \left( \ominus _1,\ominus _2 \right) =
\left( {{{a-a^3} \over {1+a^2}}} \right)^{2}.
\label{Har3}
\end{equation}
\end{mathletters}
The proof will only run if
$a \neq 1$ and $a \neq 0$, i.e.,
for entangled but not maximally entangled states
(or {\em Hardy states} \cite{CN92}).
Hardy's reasoning \cite{Hardy93} is as follows:
Consider a run of the experiment
for which $B_1$ and $B_2$ are measured and the results
``$\ominus_1$'' and ``$\ominus_2$'' are obtained.
That this will happen sometimes follows from (\ref{Har3}).
From the fact that we have ``$\ominus_1$'', it follows from
(\ref{Har1}) that if $A_2$ had been measured, we would have
obtained the result ``$-_2$''. If we assume Einstein, Podolsky,
and Rosen's (EPR) condition for elements of reality \cite{EPR35}, then
this prediction, with certainty and without disturbing the second particle,
allows us to conclude that the second particle has an element
of reality corresponding to the value ``$-_2$'' for the observable $A_2$.
By a similar argument, from property (\ref{Har2})
we conclude that the first particle has an element
of reality corresponding to the value ``$-_1$'' for the observable $A_1$.
Hence, if we had measured $A_1$ and $A_2$, instead of $B_1$ and $B_2$,
it follows from our assumptions that we would have obtained
``$-_1$'' and ``$-_2$''. However, this contradicts (\ref{Har4})
if $a \neq 1$ and $a \neq 0$.
Therefore, for a system initially prepared in a Hardy state,
the assumption that local elements of reality exist is
untenable.

\subsection*{Goldstein's version}
Goldstein's version \cite{Goldstein94} of Hardy's proof
is based on the same set of properties
of the state $\left| \psi \right\rangle$, arranged in a different way:
\begin{mathletters}
\begin{eqnarray}
P_\psi \left( -_1,-_2 \right) & = & 0,
\label{HH4} \\
P_\psi \left( {\left. \oplus_1\, \right|\, +_2} \right) & = & 1,
\label{HH2} \\
P_\psi \left( {\left. \oplus_2\, \right|\, +_1} \right) & = & 1,
\label{HH1} \\
P_\psi \left( \ominus _1,\ominus _2 \right) & = &
\left( {{{a-a^3} \over {1+a^2}}} \right)^{2}.
\label{HH3}
\end{eqnarray}
\end{mathletters}
Goldstein's reasoning is as follows:
Eq.\ (\ref{HH4}) tells us that:
(G1) Either one or both of the results of measuring $A_1$ and $A_2$
must be ``$+$''.
Eq.\ (\ref{HH2}) tells us that, if $A_2$ is ``$+_2$'', then
we can predict with certainty
and without interacting with the other spatially separated particle,
that the result ``$\oplus_1$'' will be found in a measurement of the observable $B_1$
on the first particle.
Therefore, assuming EPR elements of reality, we may conclude
that:
(G2) If $A_2$ is ``$+_2$'', then the first particle has an element of reality
corresponding to the value ``$\oplus_1$'' for the observable $B_1$.
Analogously, Eq.\ (\ref{HH1}) tells us that:
(G3) If $A_1$ is ``$+_1$'', then the second particle has an element of reality
corresponding to the value ``$\oplus_2$'' for the observable $B_2$.
It follows from (G1)-(G3) that: (G4) $B_1$ and $B_2$ cannot
simultaneously be ``$\ominus$''. However, (G4) contradicts the fact
that state $\left| \psi \right\rangle$
has, according to Eq.\ (\ref{HH3}), a nonvanishing probability for this to occur
if $a \neq 1$ and $a \neq 0$.
Therefore, for a system initially prepared in a Hardy state,
the assumption that local elements of reality exist is
untenable.

The probability of obtaining an event which contradicts local realism
is given in both versions by $P_\psi (\ominus _1,\ominus _2)$.
This probability has a maximum,
\begin{mathletters}
\begin{eqnarray}
P_{\text{max}} \left( \ominus _1,\ominus _2 \right) & = &
{\scriptstyle \left( {{{\sqrt 5-1} \over 2}} \right)^5 }, \\
 & \approx & 0.0902,
\end{eqnarray}
\end{mathletters}
for
$a={\scriptstyle \left( {{{\sqrt 5-1} \over 2}} \right)^{3/2} }$.

\section{Nonlocality without inequalities in two-particle interferometry}
In Ref.\ \cite{WXHH96} WXHH claim to have demonstrated
a violation of local realism without using inequalities for a maximally
entangled state of a bipartite two-level system.
In this Section, I will show that this is not so.

WXHH's proof uses the two-particle interferometer illustrated in Fig.\ 1.
This arrangement is a modification of the one proposed by Horne,
Shimony, and Zeilinger in Ref.\ \cite{HSZ89}. In Fig.\ 1 the source $S$ emits a pair
of particles into four beams $a$, $b$, $c$, and $d$. Each pair is in the state
\begin{equation}
\left| \zeta \right\rangle =
{1 \over {\sqrt 2}}\left( {\left| {a\,b} \right\rangle +
\left| {c\,d} \right\rangle } \right),
\label{WXHH1}
\end{equation}
where $\left| {a\,b} \right\rangle$ means particle $1$ in beam $a$
and particle $2$ in beam $b$, etc. Any experiment on particle $1$
is assumed to be spacelike separated from any experiment on
particle $2$. $M_a$ and $M_b$ are mirrors, $\phi_1$ and $\phi_2$ are
phase shifters, $BS1$, $BS2$, $BS3$, and $BS4$ are beam splitters, and
$E$, $F$, $G$, $H$, $K$, and $L$ are detectors whose efficiencies are
assumed to be 100\%.

\begin{figure}
\epsfxsize=8cm
\epsfbox{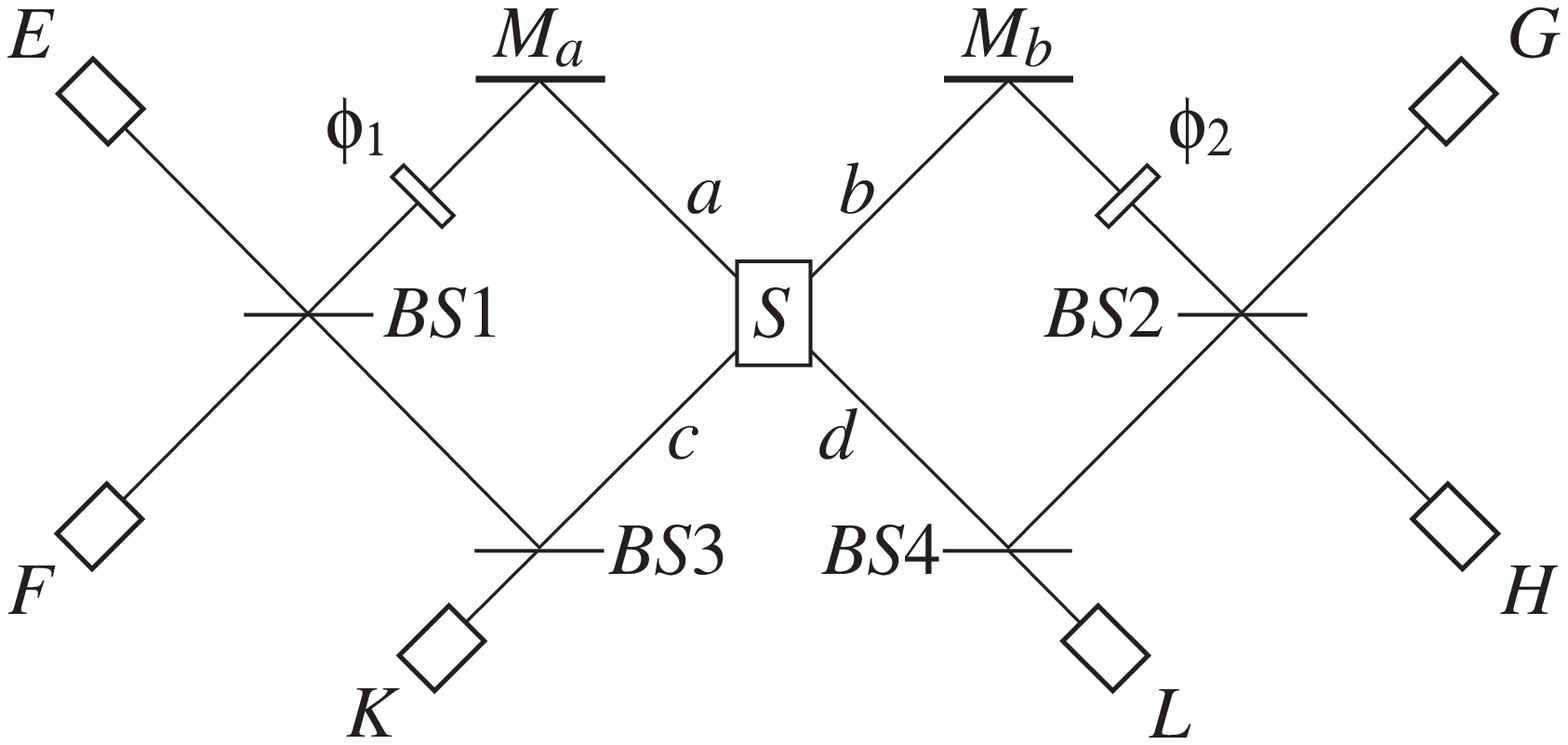}
\end{figure}
\noindent FIG.\ 1:
{\small Two-particle interferometer considered by Wu and co-workers
in Ref.\ \cite{WXHH96}.\\}

On particle $1$ we can perform one of two alternative experiments, $A_1$
and $B_1$. Each of them corresponds to the choice of the phase introduced by the
phase shifter $\phi_1$ and the reflectance and transmittance of the
beam splitter $BS1$. Similarly, on particle $2$ we can perform two
alternative experiments, $A_2$ and $B_2$, each of them corresponding to
the choice of the phase introduced by $\phi_2$ and the reflectance and
transmittance of $BS2$. WXHH choose these parameters of the experiments
$A_1$, $B_1$, $A_2$, and $B_2$ such that {\em if one selects those runs
of the experiments for which particle $1$ does not end in detector $K$, while
in the same run particle $2$ does not end in detector $L$}, then, for
these selected runs,
\begin{mathletters}
\begin{eqnarray}
P \left( {\left. A_2=H\, \right|\, B_1=F} \right) & = & 1,
\label{W1} \\
P \left( {\left. A_1=F\, \right|\, B_2=H} \right) & = & 1,
\label{W2} \\
P \left( A_1=F,\, A_2=H \right) & = & 0,
\label{W3} \\
P \left( B_1=F,\, B_2=H \right) & > & 0,
\label{W4}
\end{eqnarray}
\end{mathletters}
where $P \left( {\left. A_2=H\, \right|\, B_1=F} \right)$ is the
probability of that particle $2$ being detected in $H$ when experiment
$A_2$ is performed, conditioned to the occurrence of particle
$1$ being detected in $F$ when experiment $B_2$ is performed.
It can be immediately seen that using properties (\ref{W1})-(\ref{W4})
we can develop a Hardy-like proof.
However, properties (\ref{W1})-(\ref{W4}) are not properties of the
maximally entangled state (\ref{WXHH1}) but of the state ``distilled'' after
the selection of events stated above. In Ref.\ \cite{WXHH96}, it is not
clear whether this selection of events takes place before or after the
experiments on particles $1$ and $2$. In any case, it is interesting
to realize that WXHH's conclusions do not change if the selection takes
place before the experiments on particles $1$ and $2$. In this case, the
arrangement considered by WXHH is equivalent to the two-particle interferometer
considered by Horne, Shimony, and Zeilinger \cite{HSZ89} (in which beam splitters
$BS3$ and $BS4$ are replaced by mirrors, and detectors
$K$ and $L$ are removed), assuming that the source emits pairs in state
\begin{equation}
\left| \eta \right\rangle =
P \left( {\left| {a\,b} \right\rangle + Q \left| {c\,d} \right\rangle } \right),
\;\;\; \mbox{with}\;\;\; \left| Q \right|<1,
\label{WXHH2}
\end{equation}
instead of in state (\ref{WXHH1}). However, state (\ref{WXHH2}) is
not a maximally entangled state but a Hardy state. Therefore, I conclude
that while WXHH's proof of nonlocality is correct, it is not a proof
for a maximally entangled state but for a Hardy state distilled from a
maximally entangled state. Such a distillation is always possible
by selecting a subset of events, since the degree
of entanglement of the maximally entangled state (\ref{WXHH1}) is higher than
the degree of entanglement of the Hardy state (\ref{WXHH2}).

\section{Nonlocality without inequalities using unambiguous discrimination
between non-orthogonal states}
Any attempt to extend Hardy's proof to
cover maximally entangled states
requires finding
a subset of events, all of them
referring to a maximally entangled
state, so that the correlations exhibited by such subset cannot
be reproduced by any local realistic theory.
As becomes clear after our analysis of WXHH's proof,
this subset cannot be selected
before the local measurements
involved in the proof.
Therefore, it is interesting to investigate what would happen
in a set-up in which
the selection of events necessarily occurs after the
local measurements.
A possible scenario which fulfills this requisite is
the one in which one of the
four von Neumann local measurements involved in the original proof
is replaced by a
generalized measurement which unambiguously discriminate between
non-orthogonal states \cite{Ivanovic87,Dieks87,Peres88,Peres93}.
This scenario was
considered by Chefles and Barnett for a
different purpose: restoring local realism in the Goldstein's version
of Hardy's proof \cite{CB97}.
In the following, I will demonstrate that it is impossible to develop
a proof of nonlocality in this scenario, except in the
particular case considered by Hardy, in which
the generalized measurement discriminates between orthogonal states.

Consider the state $\left| \psi \right\rangle$ defined in Eq.\ (\ref{H1}),
and the following change of basis for the states of the first particle:
\begin{mathletters}
\begin{eqnarray}
\left| {\hat +} \right\rangle _1 & = &
\cos \alpha \left| + \right\rangle _1+
\sin \alpha \left| - \right\rangle _1, \\
\left| {\hat -} \right\rangle _1 & = &
-\sin \beta \left| + \right\rangle _1+
\cos \beta \left| - \right\rangle _1,
\end{eqnarray}
\end{mathletters}
being
$\alpha -\beta \ne \left( {\frac{1}{2}+n } \right) \pi$,
with $n$ integer.
Note that
$\left\{ \left| {\hat +} \right\rangle_1,
\left| {\hat -} \right\rangle_1 \right\}$
is a non-orthogonal basis since
\begin{equation}
\left\langle {\hat -} \mathrel{\left | {\vphantom {- +}} \right.
\kern-\nulldelimiterspace} {\hat +} \right\rangle _1=
\sin \left( {\alpha -\beta } \right).
\label{alfa}
\end{equation}
Consider the following change of basis for the second particle:
\begin{mathletters}
\begin{eqnarray}
\left| {\hat +} \right\rangle _2 & = &
\cos \gamma \left| + \right\rangle _2+
\sin \gamma \left| - \right\rangle _2, \\
\left| {\hat -} \right\rangle _2 & = &
-\sin \gamma \left| + \right\rangle _2+
\cos \gamma \left| - \right\rangle _2.
\end{eqnarray}
\end{mathletters}
These changes of basis are illustrated in {\mbox Fig.\ 2}.
The inverse transformations are:
\begin{mathletters}
\begin{eqnarray}
\left| + \right\rangle _1 & = &
M\left( {\cos \beta \left| {\hat +} \right\rangle _1-
\sin \alpha \left| {\hat -} \right\rangle _1} \right),
\label{tr1} \\
\left| - \right\rangle _1 & = &
M\left( {\sin \beta \left| {\hat +} \right\rangle _1+
\cos \alpha \left| {\hat -} \right\rangle _1} \right),
\label{tr2}
\end{eqnarray}
\end{mathletters}
where
\begin{equation}
M=\sec \left( {\alpha-\beta } \right),
\label{M}
\end{equation}
and
\begin{mathletters}
\begin{eqnarray}
\left| + \right\rangle _2 & = &
\cos \gamma \left| {\hat +} \right\rangle _2-
\sin \gamma \left| {\hat -} \right\rangle _2,
\label{tr3} \\
\left| - \right\rangle _2 & = &
\sin \gamma \left| {\hat +} \right\rangle _2+
\cos \gamma \left| {\hat -} \right\rangle _2.
\label{tr4}
\end{eqnarray}
\end{mathletters}
$\left\{ \left| {\hat +} \right\rangle_2,
\left| {\hat -} \right\rangle_2 \right\}$
is an orthonormal basis for the second particle.
With the changes of basis given by
Eqs.\ (\ref{tr1}), (\ref{tr2}), (\ref{tr3}), (\ref{tr4}),
and choosing $\gamma$ such that
\begin{equation}
\cot \gamma =\sqrt{2} \cot \theta -\cot \alpha,
\label{beta}
\end{equation}
the state given by Eq.\ (\ref{H1}) can be rewritten as
\begin{equation}
\left| \psi \right\rangle =
M\left( {q\left| {\hat +\hat +} \right\rangle +
r\left| {\hat +\hat -} \right\rangle +
s\left| {\hat -\hat +} \right\rangle } \right),
\label{adan1}
\end{equation}
where
\begin{eqnarray}
q & = & a \cos \beta \cos \gamma + b \sin(\beta+\gamma),
\label{q} \\
r & = & -a \cos \beta \sin \gamma + b \cos(\beta+\gamma),
\label{r} \\
s & = & -a \sin \alpha \cos \gamma + b \cos(\alpha+\gamma).
\label{s}
\end{eqnarray}

\begin{figure}
\begin{picture}(100,130)(-51,0){\begin{picture}(100,150)(0,0)
\put(50,10){\vector(1,0){100}}
\put(155,8){$\left| {+} \right\rangle_1$}
\put(50,10){\vector(0,1){100}}
\put(42,115){$\left| {-} \right\rangle_1$}
\put(50,10){\vector(4,1){97}}
\put(151,36){$\left| {\hat +} \right\rangle_1$}
\put(50,10){\vector(-2,3){55.5}}
\put(-15,99){$\left| {\hat -} \right\rangle_1$}
\put(50,10){\vector(2,3){55.5}}
\put(109,97){$\left| {\hat \oplus} \right\rangle_1$}
\put(50,10){\vector(-3,2){83.2}}
\put(-50,75){$\left| {\hat \ominus} \right\rangle_1$}
\put(114,16){$\alpha$}
\put(76,32){$\delta - \alpha$}
\put(33,52){$\beta$}
\put(-17,59){$\delta - \beta$}
\end{picture}}
{\begin{picture}(100,150)(103,140)
\put(50,10){\vector(1,0){100}}
\put(155,8){$\left| {+} \right\rangle_2$}
\put(50,10){\vector(0,1){100}}
\put(42,115){$\left| {-} \right\rangle_2$}
\put(50,10){\vector(2,1){89.4}}
\put(145,57){$\left| {\hat +} \right\rangle_2$}
\put(50,10){\vector(-1,2){44.7}}
\put(-7,108){$\left| {\hat -} \right\rangle_2$}
\put(50,10){\vector(2,3){55.5}}
\put(109,97){$\left| {\hat \oplus} \right\rangle_2$}
\put(50,10){\vector(-3,2){83.2}}
\put(-50,75){$\left| {\hat \ominus} \right\rangle_2$}
\put(95,19){$\gamma$}
\put(78,34){$\epsilon$}
\put(37,49){$\gamma$}
\put(20,38){$\epsilon$}
\end{picture}}
\end{picture}
\end{figure}
\vspace{5cm}
\noindent FIG.\ 2:
{\small Different basis for the first (up) and
the second particle (down) used in Sec.\ IV.\\}

Now consider an additional change of basis
for the first particle:
\begin{mathletters}
\begin{eqnarray}
\left| {\hat \oplus } \right\rangle _1 & = &
M \left[ {\cos\left(\delta-\beta\right)\left| {\hat +} \right\rangle _1+
\sin\left(\delta-\alpha\right)\left| {\hat -} \right\rangle _1} \right], \\
\left| {\hat \ominus } \right\rangle _1 & = &
M \left[ {-\sin\left(\delta-\beta\right)\left| {\hat +} \right\rangle _1+
\cos\left(\delta-\alpha\right)\left| {\hat -} \right\rangle _1} \right].
\end{eqnarray}
\end{mathletters}
$\left\{ \left| {\hat \oplus} \right\rangle_1,
\left| {\hat \ominus} \right\rangle_1 \right\}$ is an orthonormal basis.
The relation between this basis and the previous one is shown in Fig.\ 2.
The inverse transformations are:
\begin{mathletters}
\begin{eqnarray}
\left| {\hat + } \right\rangle _1 & = &
{\cos\left(\delta-\alpha\right)\left| {\hat \oplus} \right\rangle _1
-\sin\left(\delta-\alpha\right)\left| {\hat \ominus} \right\rangle _1}, \\
\left| {\hat - } \right\rangle _1 & = &
{\sin\left(\delta-\beta\right)\left| {\hat \oplus} \right\rangle _1+
\cos\left(\delta-\beta\right)\left| {\hat \ominus} \right\rangle _1}.
\end{eqnarray}
\end{mathletters}
In this new basis, and choosing $\delta$ such that,
\begin{equation}
\tan \delta ={{\sin \alpha +{\textstyle{q \over s}}\cos \beta } \over
{\cos \alpha -{\textstyle{q \over s}}\sin \beta }}\,,
\label{delta}
\end{equation}
the state $\left| \psi \right\rangle$
has the form
\begin{eqnarray}
\left| \psi \right\rangle & = &
M\left\{\left[ q \cos \left( {\delta-\alpha} \right) +
s \sin \left( {\delta-\beta} \right) \right]
\left| {\hat \oplus \hat +} \right\rangle \right. \nonumber \\
& & \left. {+\,r \cos \left( {\delta-\alpha} \right) \left| {\hat \oplus \hat -} \right\rangle
-r \sin \left( {\delta-\alpha} \right) \left| {\hat \ominus \hat -}\right\rangle } \right\}.
\label{adan2}
\end{eqnarray}

Now consider an additional change of basis
for the second particle:
\begin{mathletters}
\begin{eqnarray}
\left| {\hat \oplus } \right\rangle _2 & = &
\cos\epsilon\left| {\hat +} \right\rangle _2+
\sin\epsilon\left| {\hat -} \right\rangle _2,\\
\left| {\hat \ominus } \right\rangle _2 & = &
-\sin\epsilon\left| {\hat +} \right\rangle _2+
\cos\epsilon\left| {\hat -} \right\rangle _2.
\end{eqnarray}
\end{mathletters}
$\left\{ \left| {\hat \oplus} \right\rangle_2,
\left| {\hat \ominus} \right\rangle_2 \right\}$ is an orthonormal basis,
as illustrated in Fig.\ 2. The inverse transformations are:
\begin{mathletters}
\begin{eqnarray}
\left| {\hat + } \right\rangle _2 & = &
\cos\epsilon\left| {\hat \oplus} \right\rangle _2-
\sin\epsilon\left| {\hat \ominus} \right\rangle _2, \\
\left| {\hat - } \right\rangle _2 & = &
\sin\epsilon\left| {\hat \oplus} \right\rangle _2+
\cos\epsilon\left| {\hat \ominus} \right\rangle _2.
\end{eqnarray}
\end{mathletters}
In this new basis, and choosing $\epsilon$ such that
\begin{equation}
\tan \epsilon = {r \over q},
\label{epsilon}
\end{equation}
the state $\left| \psi \right\rangle$
has the form
\begin{eqnarray}
\left| \psi \right\rangle & = &
M\left[\left( q \cos\epsilon + r \sin\epsilon \right)
\left| {\hat{+} \hat{\oplus}}\right\rangle
+ s \cos\epsilon\left| {\hat{-} \hat{\oplus}} \right\rangle \right. \nonumber \\
& & \left. {-\,s \sin\epsilon\left| {\hat{-} \hat{\ominus}} \right\rangle } \right].
\label{adan3}
\end{eqnarray}

In addition, as can be easily checked,
\begin{mathletters}
\begin{eqnarray}
P_\psi \left( {\hat \ominus _1,\hat \ominus _2} \right)
 & = & \left[ {M s \sin\epsilon \cos\left(\delta-\beta\right)} \right]^2,
\label{mm1} \\
 & = & \left[ {M r \cos\epsilon \cos\left(\delta-\alpha\right)} \right]^2.
\label{mm2}
\end{eqnarray}
\end{mathletters}
$P_\psi \left( {\hat \ominus _1,\hat \ominus _2} \right) $
is only a function of $\theta$ (the angle that
characterizes the degree of entanglement of the
state we are considering), and
$\alpha$ and $\beta$ (the angles that characterize the
type of basis ---orthogonal or not--- we are
using to describe the state of the first particle).
The angles $\gamma$, $\delta$, and $\epsilon$
are fixed by, respectively, Eqs.\ (\ref{beta}), (\ref{delta}),
and (\ref{epsilon}).

If $\alpha - \beta = n \pi$, with $n$ integer, the scalar
product in Eq.\ (\ref{alfa}) vanishes, and then we recover a standard Hardy's
proof using orthogonal basis for each particle. In particular,
if $\alpha = \beta = 0$, then
$P_\psi \left( {\hat \ominus _1,\hat \ominus _2} \right)$ gives
the probability of obtaining an event which contradicts local
realism given by Eqs.\ (\ref{Har3}) and (\ref{HH3}).

Now let me introduce some notations:
Let $\hat{A}_2$ be the von Neumann measurement
to discriminate between the orthogonal states of the second particle
$\left| {\hat + } \right\rangle _2$ and $\left| {\hat - } \right\rangle _2$.
The only possible results of measuring
$\hat{A}_2$ are ``$\hat{+}_2$'' and
``$\hat{-}_2$''. Analogously,
let $\hat{B}_1$ ($\hat{B}_2$) be the von Neumann measurement
which discriminates between the orthogonal states of particle $1$ ($2$)
$\left| {\hat \oplus } \right\rangle _1$
($\left| {\hat \oplus } \right\rangle _2$)
and $\left| {\hat \ominus } \right\rangle _1$
($\left| {\hat \ominus } \right\rangle _2$).
On the other hand, the states
$\left| {\hat + } \right\rangle _1$ and $\left| {\hat - } \right\rangle _1$
are not orthogonal. To unambiguously discriminate between them, we define a
positive operator valued
measure \cite{Ivanovic87,Dieks87,Peres88,Peres93}, $\hat{A}_1$.
Then, the possible results of measuring $\hat{A}_1$ are
``$\hat{+}_1$'', ``$\hat{-}_1$'', or an inconclusive result ``$\hat{?}_1$''.

Hardy's proof is based on four incompatible experiments.
As seen in Sec.\ II, three of them are used to make predictions with
certainty, to define,
via EPR's condition, certain elements of reality
that cannot be reconciled with {\em some}
results of the fourth experiment. Then the proof only applies to some
runs of the fourth experiment.
In the following, I will refer to those events
as ``events for which local realism leads to a contradiction''.
On the other hand, the presence of a generalized measurement
introduces a new element in our analysis. In particular,
the possibility of an inconclusive result implies that
Hardy's (or Goldstein's) reasoning cannot be applied to
a certain subset of events. I will refer to those
events as ``events for which
the proof cannot be applied to''. In fact,
these subsets of events are different in
Hardy's and Goldstein's versions of the proof.

\subsection*{Hardy-like reasoning}
{\em If one selects all runs of the experiment except those
in which
the result of measuring $\hat{A}_1$ is inconclusive and the result of
measuring $\hat{B}_2$ is ``$\hat{\ominus}_2$''},
then, for these selected runs,
\begin{mathletters}
\begin{eqnarray}
P \left( {\left. \hat{-}_2\, \right|\, \hat \ominus_1} \right) & = & 1,
\label{HarG1} \\
P \left( {\left. \hat{-}_1\, \right|\, \hat \ominus_2} \right) & = & 1,
\label{HarG2} \\
P \left( \hat{-}_1,\hat{-}_2 \right) & = & 0,
\label{HarG4} \\
P \left( \hat \ominus _1, \hat \ominus _2 \right) & > & 0.
\label{HarG3}
\end{eqnarray}
\end{mathletters}
Property (\ref{HarG3}) only occurs por certain combinations of
$\theta$, $\alpha$, and $\beta$.
Therefore, for these selected runs a Hardy-like reasoning
like the one in Sec.\ II can be applied.
Hardy's reasoning cannot be applied
to those events in which
the result of measuring $\hat{A}_1$ is inconclusive and the result of
measuring $\hat{B}_2$ is ``$\hat{\ominus}_2$''.
Note, however, that Hardy's reasoning still applies if
the result of measuring $\hat{A}_1$ is inconclusive and the result of
measuring $\hat{B}_2$ is ``$\hat{\oplus}_2$''.

\subsection*{Goldstein-like reasoning}
{\em If one selects all runs of the experiment except those
in which the result of measuring
$\hat{A}_1$ is inconclusive and the result of measuring $\hat{A}_2$ is
``$\hat{-}_2$''}, then, for these selected runs,
\begin{mathletters}
\begin{eqnarray}
P \left( \hat{-}_1,\hat{-}_2 \right) & = & 0,
\label{HHG4} \\
P \left( {\left. \hat \oplus_1\, \right|\, \hat{+}_2} \right) & = & 1,
\label{HHG2} \\
P \left( {\left. \hat \oplus_2\, \right|\, \hat{+}_1} \right) & = & 1,
\label{HHG1} \\
P \left( \hat \ominus _1,\hat \ominus _2 \right) & > & 0.
\label{HHG3}
\end{eqnarray}
\end{mathletters}
Property (\ref{HHG3}) only occurs por certain combinations of
$\theta$, $\alpha$, and $\beta$.
Therefore, for these selected runs a Goldstein-like reasoning
like the one in Sec.\ II can be applied.
Goldstein's reasoning cannot be applied
to those events in which the result of measuring
$\hat{A}_1$ is inconclusive and the result of measuring $\hat{A}_2$ is
``$\hat{-}_2$''.
Note that Goldstein's reasoning still goes through if
the result of
$\hat{A}_1$ is inconclusive and the result of $\hat{A}_2$ is
``$\hat{+}_2$''.

\subsection*{Discussion}
In contrast to WXHH's set up, in the scenario examined
in this Section, the selection of events can
only take place after the local experiments on particles $1$ and
$2$. This raises the new problem of whether this postselection is
legitimate in a proof of nonlocality.
The only way
to develop such proof, without making any additional assumptions,
would be to show that, considering all runs of the experiment, the
probability of obtaining an event for which
local realism leads to a
contradiction using a Hardy-like (or a Goldstein-like) reasoning
is greater than the probability of
obtaining an event for which the reasoning cannot be applied.
In both versions of the proof, the probability of obtaining an event for which
local realism leads to a
contradiction is $P_\psi \left( \hat{\ominus}_1,\hat{\ominus}_2 \right)$.
However, the probability of finding an event which
the proof cannot be applied to is different for each version.
Hardy's reasoning cannot be applied
to those events in which
the result of measuring $\hat{A}_1$ is inconclusive and the result of
measuring $\hat{B}_2$ is ``$\hat{\ominus}_2$''.
Therefore, we can prove the impossibility of
local realism using Hardy's reasoning if
\begin{equation}
P_\psi \left( \hat{\ominus}_1,\hat{\ominus}_2 \right) >
P_\psi \left( \hat{?}_1,\hat{\ominus}_2 \right).
\label{Hardyyep}
\end{equation}

On the other hand, Goldstein's reasoning cannot be applied
to those events in which the result of measuring
$\hat{A}_1$ is inconclusive and the result of measuring $\hat{A}_2$ is
``$\hat{-}_2$''.
Therefore, we can prove the impossibility of
local realism using Goldstein's reasoning if
\begin{equation}
P_\psi \left( \hat{\ominus}_1,\hat{\ominus}_2 \right) >
P_\psi \left( \hat{?}_1,\hat{-}_2 \right).
\label{Goldsteinyep}
\end{equation}

Therefore, to elucidate whether a proof of
nonlocality is possible, we have to find out whether Eqs.\ (\ref{Hardyyep})
and (\ref{Goldsteinyep}) are satisfied. For this
purpose we shall use the result obtained for
$P_\psi \left( \hat{\ominus}_1,\hat{\ominus}_2 \right)$
in Eq.\ (\ref{mm1}) or Eq.\ (\ref{mm2}).
On the other hand, $P_\psi \left( \hat{?}_1,\hat{-}_2 \right)$
can be calculated \cite{Peres93} as
\begin{equation}
P_\psi \left( \hat{?}_1,\hat{-}_2 \right)=
{\rm Tr} \left[ {\left( {\hat{O}_1\otimes \left| {\hat -}
\right\rangle _2\left\langle {\hat -} \right|_2}
\right)\left| \psi \right\rangle \left\langle \psi \right|}
\right],
\end{equation}
where
\begin{equation}
\hat{O}_1=1\;\!\!\!\mbox{l}-{{2 1\;\!\!\!\mbox{l}-\left| {\hat +}
\right\rangle _1\left\langle {\hat +} \right|_1-\left| {\hat -}
\right\rangle _1\left\langle {\hat -} \right|_1} \over {1+\left|
{\left\langle {{\hat -}} \mathrel{\left | {\vphantom {{\hat -}
{\hat +}}} \right. \kern-\nulldelimiterspace} {{\hat +}}
\right\rangle _1} \right|}},
\label{PO}
\end{equation}
is a positive operator associated with the inconclusive answer
which belongs to a positive operator valued measure \cite{Peres93}.
An alternative way to calculate
$P_\psi \left( \hat{?}_1,\hat{-}_2 \right)$
is:
\begin{equation}
P_\psi \left(\hat{?} _1,\hat{-}_2 \right)=
P_\psi \left(\hat -_2\right)-P_\psi \left(\hat +_1,\hat -_2\right)
-P_\psi \left(\hat -_1,\hat -_2\right),
\end{equation}
where $P_\psi \left(\hat -_1,\hat -_2\right)$ is zero according to
Eq.\ (\ref{adan1}), and
\begin{mathletters}
\begin{eqnarray}
P_\psi \left(\hat -_2\right) & = & P_\psi \left(+_1,\hat -_2\right)+
P_\psi \left(-_1,\hat -_2\right), \\
 & = & \left( {-a\sin \gamma +b\cos \gamma } \right)^2+
\left( {b\sin \gamma } \right)^2.
\end{eqnarray}
\end{mathletters}
$P_\psi \left(\hat +_1,\hat -_2\right)$ is the probability to
unambiguously discriminate between the states
$\left| {\hat +} \right\rangle _1$ and
$\left| {\hat -} \right\rangle _1$
(given by
$1-\left| {\left\langle {{\hat -}}
\mathrel{\left | {\vphantom {{\hat -} {\hat +}}} \right.
\kern-\nulldelimiterspace} {{\hat +}} \right\rangle _1} \right|$), times
the probability to obtain ``$\hat +_1,\,\hat -_2$'' when
the discrimination succeeds,
\begin{equation}
P_\psi \left(\hat +_1,\hat -_2\right) =
{{\left[ {1-\left| {\sin \left( {\alpha -\beta } \right)} \right|} \right]
\left( Mr \right)^2}}.
\label{29}
\end{equation}

Analogously, $P_\psi \left( \hat{?}_1,\hat{\ominus}_2 \right)$
can be calculated as
\begin{equation}
P_\psi \left( \hat{?}_1,\hat{\ominus}_2 \right)=
{\rm Tr} \left[ {\left( {\hat{O}_1\otimes \left| {\hat \ominus}
\right\rangle _2\left\langle {\hat \ominus} \right|_2}
\right)\left| \psi \right\rangle \left\langle \psi \right|}
\right],
\end{equation}
where $\hat{O}_1$ is the positive operator defined in Eq.\ (\ref{PO}).
As before, an alternative way to calculate
$P_\psi \left( \hat{?}_1,\hat{\ominus}_2 \right)$
would be as follows:
\begin{equation}
P_\psi \left( \hat{?} _1,\hat \ominus_2 \right)=
P_\psi \left(\hat \ominus_2 \right)-P_\psi \left(\hat +_1,\hat{\ominus}_2\right)
-P_\psi \left(\hat -_1,\hat{\ominus}_2\right),
\end{equation}
where $P_\psi \left(\hat +_1,\hat{\ominus}_2 \right)$ is zero according to
Eq.\ (\ref{adan3}), and
\begin{mathletters}
\begin{eqnarray}
P_\psi \left(\hat{\ominus}_2 \right) & = & P_\psi \left(+_1,\hat{\ominus}_2\right)+
P_\psi \left(-_1,\hat{\ominus}_2 \right), \\
 & = & \left[ {-a\sin \left(\gamma+\epsilon \right) +
b\cos \left(\gamma+\epsilon \right)} \right]^2 \nonumber\\
 & & + \left[ {b\sin \left(\gamma+\epsilon \right) } \right]^2.
\end{eqnarray}
\end{mathletters}
$P_\psi \left(\hat -_1,\hat{\ominus}_2 \right)$ is the probability to
unambiguously discriminate between the states
$\left| {\hat +} \right\rangle _1$ and
$\left| {\hat -} \right\rangle _1$, times
the probability to obtain ``$\hat -_1,\,\hat{\ominus}_2$'' when
the discrimination succeeds,
\begin{equation}
P_\psi \left(\hat -_1,\hat{\ominus}_2 \right) =
{{\left[ {1-\left| {\sin \left( {\alpha -\beta } \right)} \right|} \right]
\left( Ms \sin \epsilon \right)^2}}.
\label{29b}
\end{equation}
As can be checked, in the limit in which we
recover Hardy's proof (i.e., if
$\alpha - \beta = n \pi$, with $n$ integer)
both $P_\psi \left( \hat{?} _1,\hat -_2 \right)$ and
$P_\psi \left( \hat{?} _1,\hat \ominus_2 \right)$ are zero.
However, a detailed numerical examination reveals that
for every $\theta$, $\alpha$ or $\beta$, Eqs.\ (\ref{Hardyyep}) and
(\ref{Goldsteinyep}) are never satisfied. Therefore,
assuming that the argument developed in this
section is the most comprehensive
based on the idea of replacing a
von Neumann measurement with a measurement which discriminates between
non-orthogonal states,
I conclude that no proof of Bell's theorem
without inequalities based on
such idea can work, except in the
particular case considered by Hardy, in which
the generalized measurement discriminates between orthogonal states.
In particular, no proof of nonlocality without inequalities for
maximally entangled states of bipartite two-level systems can be
developed in this scenario.

\section{Conclusions}
There is a proof of nonlocality without inequalities for
bipartite {\em three-level} maximally entangled
states \cite{Cabello98}.
However, so far, no attempt to extend Hardy's proof to bipartite two-level
maximally entangled states works.
In the proof by Wu and co-workers \cite{WXHH96}, the source emits a
maximally entangled state. However, the state after the selection
is, as in Hardy's proof, entangled but nonmaximally entangled.
On the other hand,
it has been proved
that it is impossible to generalize Hardy's proof by replacing
one of the four von Neumann measurements
with a measurement to unambiguously discriminate between
non-orthogonal states. Therefore,
neither this scenario can be used to extend the proof
to maximally entangled states.\\

\section*{Acknowledgments}
The author thanks Jos\'{e} L. Cereceda
for his contributions and suggestions to this work.
The author also thanks Stephen Barnett, Anthony Chefles,
Gonzalo Garc\'{\i}a de Polavieja, Carlos Serra,
and Guifr\'{e} Vidal for their valuable comments.
This work was financially supported
by the Universidad de Sevilla (Grant No.\ OGICYT-191-97)
and the Junta de Andaluc\'{\i}a (Grant No.\ FQM-239).

\end{document}